\documentclass[manuscript]{acmart} %anonymous
\usepackage{float}
\usepackage{multirow}
\usepackage{booktabs}
\usepackage{array}
\usepackage{url}
\AtBeginDocument{%
  }

\definecolor{darkgreen}{rgb}{0,0.5,0}
\definecolor{orange}{rgb}{1,0.5,0}
\definecolor{teal}{rgb}{0,0.5,0.5}
\definecolor{darkpurple}{rgb}{0.5, 0, 0.5}
\definecolor{olive}{rgb}{0.6,0.6,0}

\begin{document}

%%
%% The "title" command has an optional parameter,
%% allowing the author to define a "short title" to be used in page headers.
\title{Breaking the Midas Spell:\\
Understanding Progressive Novice-AI Collaboration in Spatial Design
}
% \title{Wizards-mediated Appropriation:\\
% Understanding Novice-AI Collaboration in Progressive Spatial Design
% }
% \title{Wizards-mediated Workflows:\\
% Understanding Novice-AI Collaboration in Progressive Spatial Design
% }

%%
%% The "author" command and its associated commands are used to define
%% the authors and their affiliations.
%% Of note is the shared affiliation of the first two authors, and the
%% "authornote" and "authornotemark" commands
%% used to denote shared contribution to the research.
\author{Zijun Wan}
\authornotemark[1]
\affiliation{%
  \institution{Bartlett School of Architecture}
  \city{London}
  \country{United Kingdom}}
\email{ucbvzw6@ucl.ac.uk}

\author{Jiawei Tang}
\affiliation{%
  \institution{Beijing Institute of Technology}
  \city{Beijing}
  \country{China}}
\email{tangjiawei815@gmail.com}

\author{Linghang Cai}
\affiliation{%
  \institution{ArtCenter College of Design}
  \city{Pasaden}
  \country{United States}}
\email{cailinghang17@gmail.com}

\author{Xin Tong}
\affiliation{%
  \institution{Hong Kong University of Science and Technology (Guangzhou)}
  \city{Guangzhou}
  \country{China}
}
\email{txmaylnjz@gmail.com}

\author{Can Liu}
\affiliation{%
  \institution{School of Creative Media, City University of Hong Kong}
  \city{Hong Kong}
  \country{China}}
\email{can.liu.eu@gmail.com}
%%
%% By default, the full list of authors will be used in the page
%% headers. Often, this list is too long, and will overlap
%% other information printed in the page headers. This command allows
%% the author to define a more concise list
%% of authors' names for this purpose.
%\renewcommand{\shortauthors}{Trovato et al.}

%%
%% The abstract is a short summary of the work to be presented in the
%% article.
\begin{abstract}
  %In spatial design, Artificial Intelligence (AI) integration has predominantly targeted professionals, requiring users to have substantial prior knowledge and expertise. \cl{I don't think AI tools are only made for professionals, I would remove this motivation.} 
  In spatial design, Artificial Intelligence (AI) tools often generate the entire spatial design outcome in a single automated step, rather than engaging users in a deepening and iterative process. This significantly reduces users' involvement, learning, and creative capabilities, leading to a superficial understanding of spatial design. We conducted a Wizard-of-Oz study, where Novices and AI (acted by experimenters) worked together to finish spatial design tasks using various AI models. %Then, a pattern mapping and a thematic analysis were conducted to 
  We identified typical function and workflow patterns adopted by the participants, leading to the understanding of the opportunities and challenges in the human-AI co-creation process. Based on insights gathered from this research, we proposed some design implications of the novice-AI collaboration system that aims to democratize spatial design through a progressive, iterative co-creation process. 
\end{abstract}

%%
%% The code below is generated by the tool at http://dl.acm.org/ccs.cfm.
%% Please copy and paste the code instead of the example below.
%%


%%
%% Keywords. The author(s) should pick words that accurately describe
%% the work being presented. Separate the keywords with commas.
%% A "teaser" image appears between the author and affiliation
%% information and the body of the document, and typically spans the
%% page.
\begin{teaserfigure}
  \includegraphics[width=\textwidth]{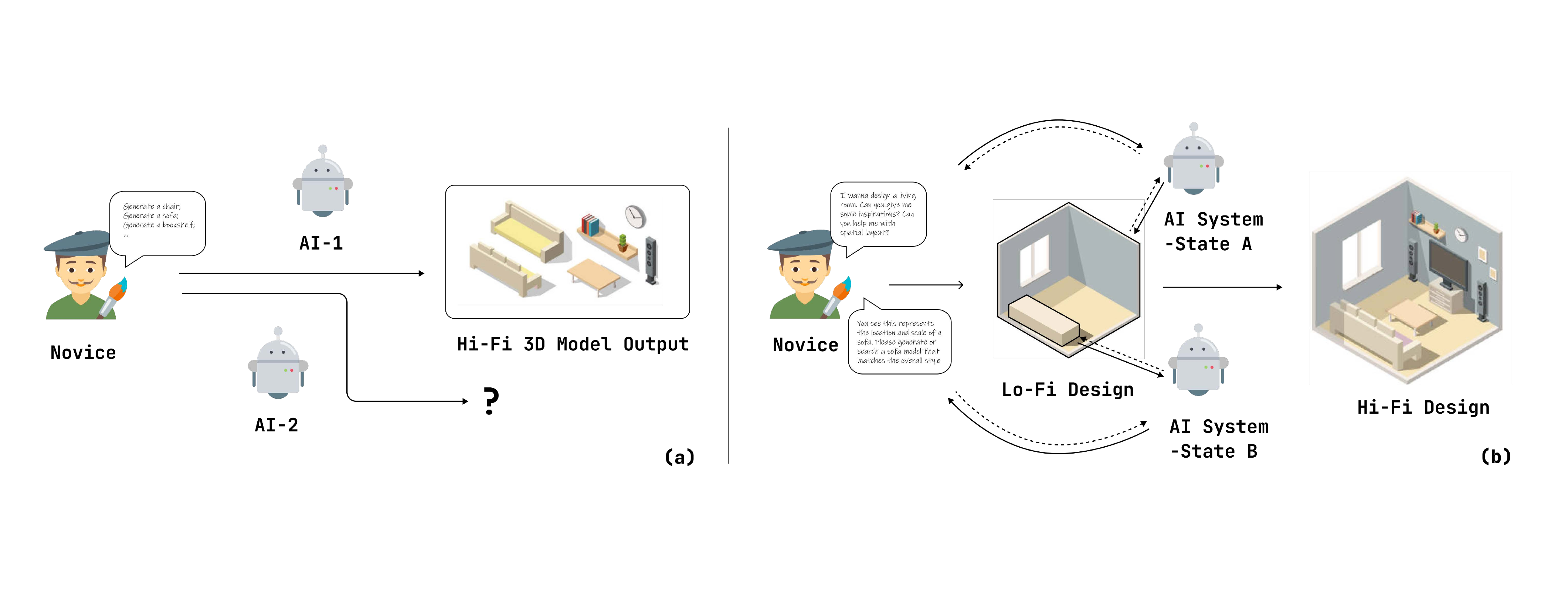}
  \caption{(a) the pipeline of the traditional AI-driven tools generates the output in one step, and these tools are scattered; (b) an envision of a progressive human-AI collaboration framework in Spatial Design, in which AI collaborates with a novice learner in the whole progress, understanding each embodied command, helping human develop from a vague idea in mind to a low-fidelity design, and finalizing with a high-fidelity spatial design scheme.}
  \Description{Enjoying the baseball game from the third-base
  seats. Ichiro Suzuki preparing to bat.}
  \label{fig:teaser}
\end{teaserfigure}

%\received{20 February 2007}
%\received[revised]{12 March 2009}
%\received[accepted]{5 June 2009}

%%
%% This command processes the author and affiliation and title
%% information and builds the first part of the formatted document.
\maketitle

\section{Introduction}
Rapid advances in Generative AI technology have enabled AI to actively collaborate with humans in the creation of new content, particularly in writing, drawing, coding, music, and even 3D asset creation \cite{9,14,21,25,28,29}. In the field of spatial design, recent studies have highlighted AI’s potential to assist professional spatial designers in various tasks, such as rendering representations, generating spatial layouts, refining facades, etc \cite{5,16,40,44}. 
However, the effectiveness of AI in these areas still largely depends on the designers’ extensive prior knowledge and expertise. Those novices who do not receive systematic training, still face unique challenges compared to professional designers when engaging in the whole design process, for example, they often struggle to find inspiration and effectively identify and define design problems, particularly in the early stages; Additionally, they may have difficulty understanding complex spatial structures and can easily become confused by fundamental issues such as spatial scale and visual styles \cite{19,32}. 

Existing AI-assisted spatial design technologies \cite{33,46,52} are typically structured so that the user provides a prompt, and the AI generates the whole spatial design outcome in a single, automated step directly, rather than involving users in an iterative process. This restricts their learning and creative ability to contribute meaningfully, leading to a superficial understanding of design; And they are easily constrained by rigid templates or pre-set solutions, limiting the possibilities to control their design to fulfill their actual needs positively.

As references, some prototypes in creative design \cite{2,6,18,50,51} have proven effective in boosting creativity among non-experts. This allows them to gradually refine their ideas in an accessible and understandable way. There has been increasing interest in incorporating AI into these processes and implementing them, especially in areas where technologies are mature enough to implement, however, such tools are still rare in the realm of spatial design, where the complexity of space, form, and function poses unique challenges. Additionally, another significant challenge in spatial design is that it is often embodied and multimodal, which means that simply communicating with AI through text is insufficient for effective spatial design \cite{3,12}. AI needs to provide targeted in-situ responses based on the current stage and state of the design. This requires the AI to process and integrate various types of multimodal information, such as text commands provided by the user, the position and dimensions of the model, and reference images representing the stylistic choices the user may want to consider.

Building on the above gaps, the current research investigates how AI can further assist novice users in spatial design, focusing on gradual, progressive development rather than immediate high-fidelity outputs. This approach is intended to foster their active exploration, deepen their understanding of spatial design, and enhance creativity among novices, while also addressing their practical needs for detailed, personalized solutions. Specifically, we seek to explore: 
\begin{itemize}
\item {\texttt{}}What challenges are they faced with, and what help do novice users need to engage meaningfully in the whole spatial design process? 
\item {\texttt{}}What role can AI play in supporting this engagement, at each stage of spatial design?
\item {\texttt{}}What implications can we get to help design systems and tools for facilitating novice-AI collaboration?
\end{itemize}

To address these questions, we conducted a Wizard of Oz study to learn from novice AI co-creation in spatial design. This study involved organizing a Human-AI co-design session \cite{36} using the Wizard-of-Oz method \cite{8}, where the session planner acted as the AI and participants acted as users collaborating on a spatial design task. Following the task, participants completed semi-structured interviews and had their design process replayed for observation. This allowed us to gather specific user needs, identify common interaction patterns in AI collaboration, and understand opportunities and challenges in the co-creation process. From this study, we summarized important design requirements for process-oriented AI assistance tools, which fall into two main categories: 1) providing critical inspiration, prior knowledge and instructions: at various stages of the design process, particularly during concept generation and detail deepening, users require AI to provide "critical instruction", which refers to the AI’s ability to introduce novel ideas or solutions that significantly influence the direction of the design, helping users overcome creative blocks or explore new possibilities. 2) executing specific actions within an embodied context: users need AI to perform tasks such as adjusting spatial layouts, refining design elements, or correcting errors. Additionally, given the nonlinear nature of the design process [48], it also provides insights that users can have access to revisit and revise previous design iterations. This involves enabling users to browse the entire history of the project, revert to earlier versions, and modify instructions as needed. The second part of the study was informed by these findings and involved some design guidelines and  a novel interface design proposal. These guidelines respond to the above two categories of needs and users' workflows, and the initial design offers users a visualized, user-friendly AI-supported interface that guides them progressively from concept to realization. 

In summary, this research makes the following contributions:
\begin{itemize}
\item {\texttt{}}It identifies the typical interaction and workflow patterns of novice users in the spatial design process through a Wizard-of-Oz study.
\item {\texttt{}}It offers insights into how AI can be used to boost creativity, efficiency, and accuracy, as well as challenges they are facing within the whole spatial design process, through a thematic analysis of semi-structured interviews.
\item {\texttt{}}It proposes several design guidelines based on the findings in the experiment, and an intuitive interface design that understands the user's embodied commands, lists the common actions of the AI, and visualizes them as function cards. Moreover, the system is  highly adaptive, editable and traceable. 
\end{itemize}

\section{\textbf{Related Works}}

\subsection{Progressive Strategy in Spatial Design}

Design, as a creative behavior, involves a high degree of complexity, transforming abstract ideas and concepts into tangible reality. on the other hand, it also implies an ability to form value and produce innovation entailing the generation of new items \cite{42}. However, this goal cannot be achieved instantaneously \cite{48}. In creative practice, due to the ambiguity and multiplicity of design problems \cite{1}, the design process doesn’t move directly from concept to final product, instead, creators oscillate between various stages of design, making adjustments, refinements, and iterations along the way. This process is marked by the passage of time and the influx of ideas \cite{38}, during which the design evolves and matures. The Double Diamond Model \cite{53} illustrates the shifts in thinking throughout this process, where two modes of thinking—divergent and convergent—alternate multiple times. In the early stages of design, it is crucial to first identify key problems and generate a wide range of ideas. Subsequently, the process involves narrowing down and eliminating infeasible solutions, ultimately focusing on implementing viable ones. In this context, progressive Design (PD) offers an effective structured design framework to guide creators and learners in their work \cite{7}. Progressive Design is a systematic, step-by-step approach that emphasizes continuous iteration and refinement throughout the design process. This process reduces blind spots in the design process and ensures that the final product meets the intended goals through ongoing feedback and adjustments. In the field of spatial design, the principles of Progressive Design are widely applied. Designers typically begin with low-fidelity mental imagery, models, or sketches, which serve as simple visualized representations of mass and spatial relationships \cite{17}. As the design progresses, these initial models are gradually refined and enhanced in multiple iterations. The functionality, form, and details of the design are continuously improved, evolving into high-fidelity, intricate, and precise spatial designs. For novice learners, a progressive approach provides a method for deeply engaging with the design process, gradually acquiring spatial design knowledge through the feedback received during the iterative design process, and gaining the freedom to control and shape the outcome of their work \cite{31,34}. In this context, some researchers have discussed the impact of AI on the design iteration process, pointing out that AI can enhance design iteration in various ways, including creating representations, triggering empathy, and increasing engagement. However, it also highlights the current limitations of AI technology, such as its reliance on predefined patterns and rules \cite{47}.

\subsection{Human-AI Co-creation for Novice Learners}

An increasing number of studies have focused on the collaboration between humans and AI, exploring how different interaction methods impact creative outcomes and user experience. For example, a cross-modality study has demonstrated a condition-based system to dynamically match user interactions with intentions and empower different multimodal AI responses, which can significantly improve user experience \cite{15}. Other studies have identified common collaboration patterns between AI and humans \cite{13}.  In general, AI does not assist the user through every single step of the process; rather, it is called upon when needed. Throughout this process, the role of AI is continuously evolving —most commonly, it acts as an assistant, offering solutions or creative ideas, but it can also function as a mentor or overseer at times. Research has also found that users desire more proactive collaboration with AI \cite{29}, and in these situations, controllability and interpretability are key challenges to realizing this vision \cite{4}. Users want to have a certain degree of control over AI’s behavior and decisions, and they also need to clearly understand how the AI operates to ensure that collaboration remains effective and creative.  Some scholars have attempted to construct the "Human-AI Co-creation Model" from a macro perspective, which encompasses various design stages, including perceiving, thinking, expressing, collaborating, building, and testing. This model emphasizes that AI is beyond a tool, but a new philosophy, a new strategy, and a new force in the creative process \cite{42}.

Some AI applications are making efforts in the non-professional domains, where humans and AI collaborate to create new outputs. Studies focusing on non-professional users analyze their specific characteristics to design user-friendly interfaces.  For example, one study examined non-professional users with diverse backgrounds and writing styles to understand how large language models perform across different contexts \cite{20}. Other research has involved developing interface prototypes for testing and collecting user feedback. For instance, the "DeepThInk" AI-assisted digital art tool lowered technical barriers to artistic creation while enhancing users' creativity and expressive capabilities \cite{11}.  Additionally, another study on novice-AI collaboration in music creation provided beginner users with a range of user-friendly settings, allowing them to collaborate with AI to produce complex music pieces \cite{25}. However, studies reveal that much of the research in this field has focused on areas with mature technologies like painting or writing. Efforts to explore human-AI collaboration in spatial design remain relatively nascent, indicating there is significant potential for future exploration in this domain.

\subsection{AI-assisted 3D Content Creation}

In recent years, the field of AI-assisted 3D content creation has witnessed significant advancements due to the increasing volume of 3D datasets and the maturation of Gen-AI (Generative AI) technologies, offering promising potential to reduce both time and labor costs associated with 3D content creation \cite{22}. AI technologies primarily get involved in the following key areas in 3D content creation: 3D model generation, 3D retrieval, and 3D reconstruction.  

In 3D model generation, it focuses on creating 3D models directly from text, images, or 3D models’ inputs. Initially, this field relied heavily on existing 3D model datasets as training materials \cite{35}, but due to the scarcity of 3D model data, this approach was limited in its effectiveness. Recent advancements have seen rapid development in text-to-3d and image-to-3d, from early models like DreamFusion \cite{30} to more advanced systems such as One2345 \cite{23}, Stable Zero123 \cite{24}, and the latest TripoSR \cite{37}. Additionally, specialized models like ControlRoom3D \cite{33} and SceneWiz3D \cite{45} have been developed specifically for scene generation, and some of them have been used in assisting scene creation in VR/AR environments. However, generating scenes remains very challenging and immature compared to single 3D object generation, because a scene contains multiple 3D objects, requiring consideration of their spatial layout of them, as well as the stylistic coherence of the space and these objects. In terms of 3D retrieval, it involves the use of various deep learning models to search large databases for 3D models such as Objaverse \cite{10} based on input text or images, including hand-drawn sketches. Some researchers \cite{26} have demonstrated how to use sketches to assist 3D shape retrieval in VR, which offers a natural way to convey information about volumetric shapes and spatial relationships and is friendly to non-professionals. Also, 3D retrieval is considered a simple way of assisting 3D scene creation because its time cost is far less than generating new models, which is very helpful in optimizing the real-time interaction process between humans and AI \cite{43}. As for 3D reconstruction, it is particularly useful in fields such as cultural heritage preservation \cite{39}, where accurate digital replicas of physical artifacts and scenes are needed. Neural Radiance Fields (NeRF) \cite{27} have emerged as a prominent technique in this domain, offering a method to create highly detailed 3D representations from 2D images. Current commercial applications such as Luma AI and 3DPresso utilize NeRF to provide realistic 3D models, although they face challenges related to long generation times, often requiring around 30 minutes per model.

\section{A Wizard-of-Oz Study}

The aim of this “Wizard of Oz” study is to reveal what users prefer to call AI for assistance and identify common collaboration workflows,  fostering better collaboration between humans and AI agents during the spatial design process. It involves a pilot study and a formal co-design session. All participants signed informed consent forms before engaging in the formal experiment, which was approved by the ethics committee of the local University. The anticipated content to be gathered from the experiment includes: 1) What design outcomes are produced by designers during the allotted time for the design task? 2) Under what circumstances do designers seek AI assistance, and how do they typically express their need to translate into specific actions? 3) What are the typical categories of workflows for novice AI collaboration? 4) The participants' perceptions of the interactive system and process. 

\subsection{Pilot Study}

Before conducting the formal experiment, we performed a pilot study to determine the necessary background introduction, tools, and specific procedural details required for the participants of the formal co-design session. The pilot study was conducted in two phases.

\begin{figure}[h!]
  \centering
  \includegraphics[width=0.9\textwidth]{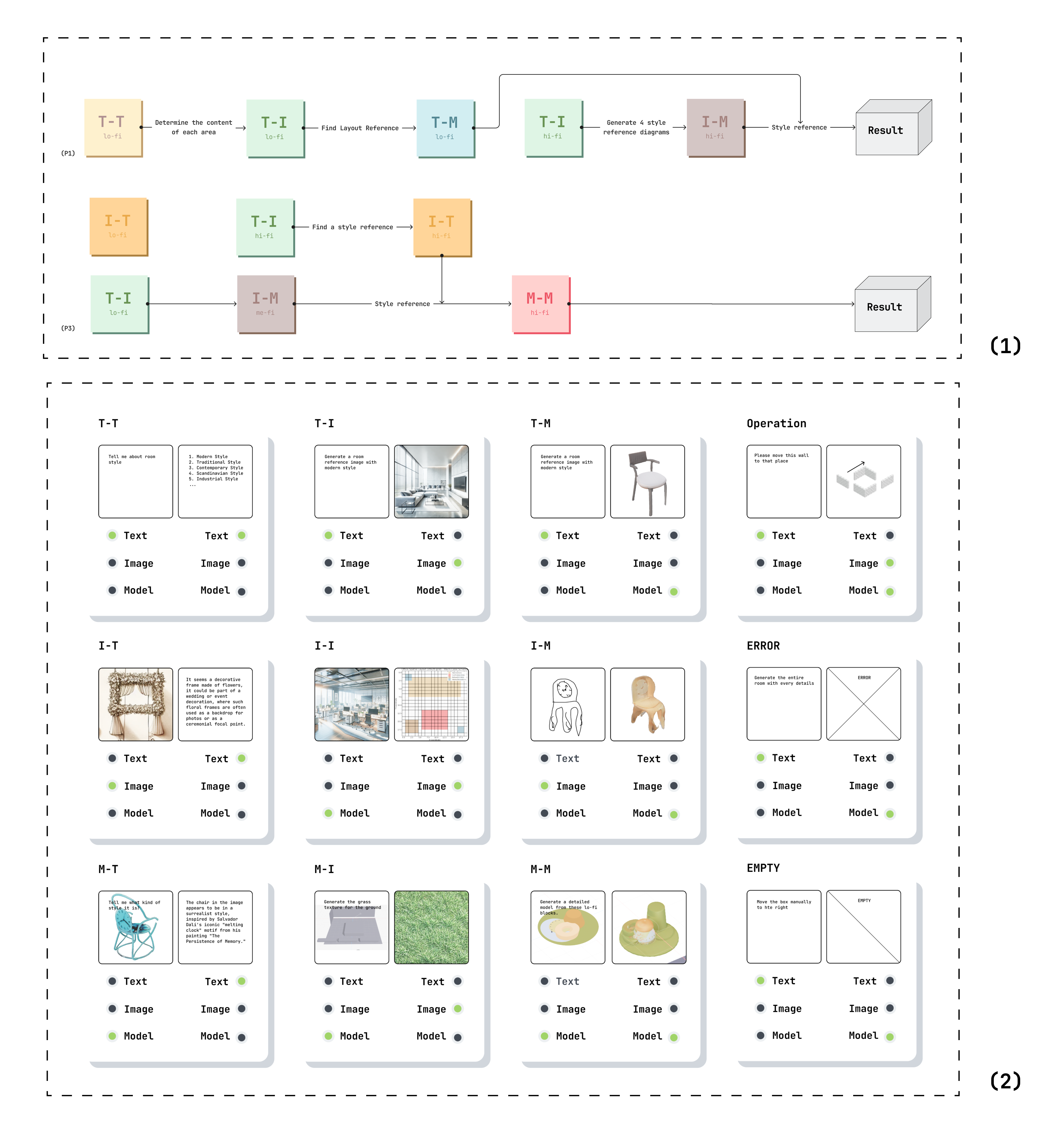}
  \caption{(1) The first pilot study with a simple sticky-note format; (2)  12 (3*3 + 1 + 2) functional card examples in the second pilot study, for example, T-I means Text to Image}
  \label{fig:pilot study}
  \Description{ (1) The first pilot study with a simple sticky-note format; (2)  12 (3*3 + 1 + 2) functional card examples in the second pilot study, for example, T-I means Text to Image}
\end{figure}

In the first pilot study, we recruited five participants through the researchers' personal networks who used simple sticky notes to conceptualize their potential workflows for using AI in progressive spatial design (Figure~\ref{fig:pilot study}.1), Using sticky notes allowed participants to freely express their thoughts without the constraints of experimental conditions or settings. This preliminary step aimed to confirm the functional classification details for the subsequent stages. there are examples of two participants’ workflows. For instance, Participant P1 was asked to design a modern-style kitchen. His approach was meticulous; he first described how he would use AI to find spatial layout references, confirm the scales of various furniture, and understand how users move through the space. Then, he would generate a reference image and attempt to model each component based on this image. In contrast, Participant P3 focused more on visual aesthetics, opting to first find a suitable visual style, which he would then apply to a low-fidelity model, either directly generated or manually created, before refining it into a high-fidelity model. Although this experiment did not display each step's real-time outcome and merely expressed the workflow as a mental map, we found that deeper analysis of such an experimental approach could reveal common workflows and frequently used functions across multiple participants. This finding led us to conduct a second pilot study.

In the second pilot study, to ensure a comprehensive understanding of common AI and non-AI operations in spatial design, our research team first engaged in internal discussions and investigations, creating 12 cards, as shown in Figure~\ref{fig:pilot study}.2: 9 of these represented the composition of inputs (text, image, model) and outputs (text, image, model), another one represented basic operation need (for example, user ask AI to make a hole on a wall).  This was included because, despite efforts to make the system user-friendly, users without modeling experience would still require some learning. Therefore, they might need AI assistance for basic operations. Additionally, there were two special cards: one represented the Empty card, indicating that AI assistance was not required, and the other was an Error card, triggered when participants made unrealistic requests to the AI (e.g., asking the AI to generate a highly detailed spatial model in one step without providing a clear description of the requirements). We then conducted a spatial design test with 2 participants, allowing them to combine these function cards with a modeling software called “Spline”. The methodology of this pilot study was the same as that of the formal experiment, both employing the "Wizard-of-Oz" method, but it was completed in one-time. The test results showed that both participants successfully completed the spatial design tasks and achieved satisfactory results. However, time was lost during the process due to the participants' exploration of the software's functionalities and the experimenter's unfamiliarity with operating the AI tools in the back-end. Therefore, after this pilot study, we held an internal discussion to refine the experimental procedure and confirm the design tasks, the time allocation for the design tasks (60 minutes), the expected outcomes, and other details.

\subsection{Wizard-Mediated Human-AI Co-design Session}
\subsubsection{Participants}
For the formal experiment, we recruited 12 novice participants in spatial design, all of these subjects were between the ages of 18-35 years old, some of the candidates did not have any experience in spatial design at all, and some others had very limited experience in spatial design.  Additionally, these non-professional participants were required to have a basic interest in spatial design to avoid a mismatch of needs.

\subsubsection{Apparatus}
The experiment will be conducted in an online format, primarily using desktop computers. The participants will mainly use Figma to read the introduction and perform card-based tasks while using Spline to create 3D spatial design models through a split-screen setup. Figma is a user-friendly collaborative tool for discussion, note-taking, and design, while Spline is a 3D modeling tool designed for non-professionals, offering only basic modeling functionalities and pre-made model components.

For the experimenter, besides Figma and Spline, they will need to open various existing AI generation tools to perform tasks requested by the participants in the back end. These tools include ChatGPT (primarily for textual conversations and image explanations), MidJourney (for image generation), ComfyUI (used for specific, controllable image workflows like style transfer and generating refined images from sketches—these workflows can be downloaded from open-source repositories on GitHub), and Meshy (for fast generation of 3D models based on cloud platforms). Additionally, to accelerate the speed of providing suitable models, we provide a manual model retrieval workflow, using Sketchfab, the largest public model library available. Based on participants’ needs, models are retrieved and imported into Spline.

\subsubsection{Experiment Design}
The experiment, which will be fully recorded and videotaped, is expected to last approximately 90 minutes. After the experimenter introduces the purpose, rules, cards, and design platforms, participants who agree to participate will sign a consent form and then proceed to watch three spatial design case studies. They will review the cards in Figma, read the descriptions and illustrations, and familiarize themselves with the basic operations and card functions in Spline. The experiment is designed in two main phases, with a rationale rooted in progressive learning and increasing task complexity. The first 30-minute design task allows participants to initially familiarize themselves with the tools and task environment, preparing for the more advanced task to follow. A 5-minute interference task \cite{54} is introduced after the first design phase to create a cognitive break. The second stage involves a more complex and higher fidelity design task, lasting another 30 minutes. This phase challenges participants to build on their initial design experience and handle more intricate details. Based on the two phases, the progressive structure of the tasks allows researchers to observe how participants improve their design capabilities when faced with increasing difficulty and how they manage detail-oriented tasks after completing the initial phase. As for the selection of questions, they are drawn from our predefined question bank, and the probability is randomized. This question bank covers various types of space design, ranging from virtual to real, and from indoor to outdoor environments, with a moderate level of difficulty.

\textbf{Task 1: Low-Fidelity Spatial Design Task (30 min).}
We will present participants with a typical scene design task. Participants will complete the design of a scene within 30 minutes, including aspects: layout, scale, and function. They will select design operation cards in Figma, including non-AI, AI-assisted, and custom options. These cards correspond to operations in the Spline platform. One researcher will monitor time and provide guidance, notifying participants when 5 minutes remain and checking if any design categories (positioning, color, function, material, lighting) are missing. another researcher acts as the AI, providing feedback based on participant requests. For example, if a participant uses a text-generation tool card, the researcher will input prompts into ChatGPT and paste the reply into Spline. If a participant selects a "Generate Model" card, the researcher will generate the model and upload it to Spline. Participants will then adjust and refine their design within Spline based on the AI feedback.

\textbf{Task 2: High-Fidelity and Reflective Spatial Design Task (30 min).}
After a 5-minute break, we invited the participants to come back and they were encouraged to review their spatial design proposals, and then refine or modify specific details. Similarly, the work in this phase focuses not only on addressing layout, scale, and function issues that may have arisen during the first phase due to time constraints or operational errors, but also places greater emphasis on detailed elements such as lighting, color, materials, and decorations.

\begin{figure}[h]
  \centering
  \includegraphics[width=\linewidth]{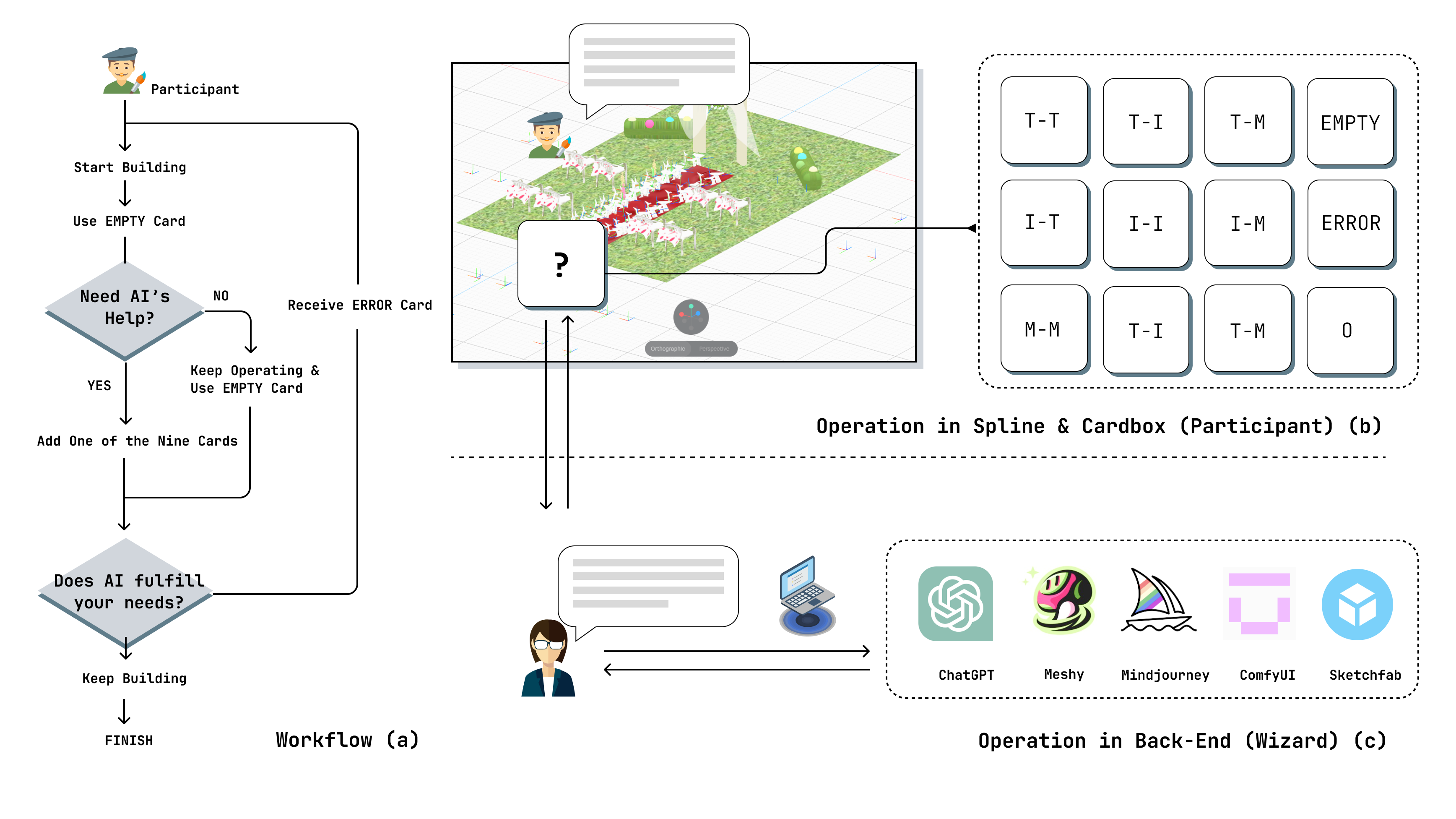}
  \caption{The formal Experiment Process: (a) the entire collaboration workflow; (b) the "front-end" operations in Spline and Cardbox, where participants can call AI for help when they are modelling in the Spline. Each call consumes one specific function card; (c) the operations in the "back-end", where wizards act as AI, give output to participants through multiple AI tools}
  \label{fig:Process}
  \Description{The formal Experiment Process}
\end{figure}

\section{Data Analysis}

After participants completed the design tasks (Task 1 and Task 2), data analysis was conducted using both quantitative and qualitative methods to provide a comprehensive understanding of their collaborative experiences with AI. The analysis consists of two parts: the first part involves “pattern mapping” of the entire interaction process between participants and AI, aiming to identify frequently used functions when calling for AI assistance and common workflows in AI-assisted spatial design. The second part is a thematic analysis of a semi-structured interview conducted with participants, providing deeper insights into the subtle nuances of their interactions with AI, their perceptions of the interaction process, and the challenges they encountered.

\subsection{Observation of Novice-AI Interaction}

During each co-design session, one of our team members used function cards developed in the pilot study phase to document participants' interactions with the AI. Each card contained information on the function (from one of the ten categories mentioned before), the specific input provided by the participant when calling for AI assistance, and the corresponding output generated by the AI. Every instance of AI interaction by participants was recorded using these cards, enabling further statistical analysis and induction. Additionally, we conducted a higher-level abstraction from the participants' interaction pattern mapping to summarize common entire workflow patterns.

\subsubsection{AI Function Usage}

We recorded the number and types of function cards used by each participant during their interactions with AI. Across the 12 participants, a total of 126 requests were made, as shown in Table~\ref{tab:Statistics}. On average, each participant asked for AI assistance around 10 times during the entire spatial design process. The most frequently used function was T-M (Text to Model), with 38 instances. This high usage can be attributed to participants' limited experience with modeling and time constraints, prompting them to request AI-generated models that directly matched their textual descriptions. In contrast, I-M (Image to Model) was used less frequently, with only 12 instances. Using images (whether sketches or reference images) was not as fast as providing direct textual commands unless participants had a clear visual requirement for the model.

\begin{table}[h]
  \caption{Statistics of AI Function Usage}
  \label{tab:Statistics}
  \centering
  \begin{tabular}{p{1cm} >{\centering\arraybackslash}p{0.75cm} >{\centering\arraybackslash}p{0.75cm} >{\centering\arraybackslash}p{0.75cm} >{\centering\arraybackslash}p{0.75cm} >{\centering\arraybackslash}p{0.75cm} >{\centering\arraybackslash}p{0.75cm} >{\centering\arraybackslash}p{0.75cm} >{\centering\arraybackslash}p{0.75cm} >{\centering\arraybackslash}p{0.75cm} >{\centering\arraybackslash}p{0.75cm} >{\centering\arraybackslash}p{2cm}}
    \toprule
    & All & T-T & T-I & T-M & I-T & I-I & I-T & M-I & M-T & M-M & O (Operation) \\
    \midrule
    All  & 126 & 10 & 26 & 38 & 10 & 1 & 12 & 0 & 0 & 9 & 19 \\
    P1   & 6   &    & 1  &    & 1  &   &  3 &   &   & 1 &    \\
    P2   & 12  & 2  & 3  & 1  & 2  &   &  4 &   &   &   & 1  \\
    P3   & 15  & 1  & 2  & 5  & 1  &   &    &   &   &   & 6  \\
    P4   & 10  & 3  & 3  & 5  & 1  &   &    &   &   &   & 1  \\
    P5   & 12  &    & 2  & 6  & 1  &   &    &   &   &   & 2  \\
    P6   & 8   &    &    & 4  &    &   &    &   &   &   & 4  \\
    P7   & 10  & 1  & 5  & 4  &    &   &    &   &   &   &    \\
    P8   & 8   &    & 2  &    &    &   &    &   &   & 6 &    \\
    P9   & 10  &    &    & 2  &    &   & 2  &   &   & 1 & 3  \\
    P10  & 12  &    & 1  &    & 1  &   & 3  &   &   &   & 1  \\
    P11  & 12  & 1  & 5  & 7  & 2  &   &    &   &   & 1 & 1  \\
    P12  & 11  & 2  &    & 4  & 1  & 1 &    &   &   &   &    \\
    \bottomrule
  \end{tabular}
\end{table}

The second most frequently used function was T-I (Text to Image), which was called 26 times. This function played a crucial role throughout the project, from generating reference images based on textual descriptions in the early stages to creating floor plans and adding detailed textures to models in later stages. On the other hand, T-T (Text to Text), which refers to human-AI dialogue, was used only 10 times. This was primarily employed in the project's initial stages when participants got inspiration from the AI. The Operation function was also widely used, with 19 requests. This suggests that participants unfamiliar with modeling still tended to rely on AI for automatic modeling operations. Some users preferred to create low-fidelity primitives (such as cubes or spheres) to represent the form, size, and location of a model and then asked the AI to generate a high-fidelity model based on those primitives. This is a common design practice among spatial design professionals, demonstrating that some participants were aware that providing such scaffolding could help the AI generate more accurate models, providing them with greater control over the final output.

Lastly, I-I (Image to Image) was used only once, while M-I (Model to Image) and M-T (Model to Text) were not used at all, indicating that these functions may not be necessary for future design iterations.

We further categorized these requests into patterns and sub-patterns, as shown in Table~\ref{tab:Categorization}, which served as the foundation for the functional types in our subsequent framework and interface design. There are primarily mainly two patterns: one is the "Guider," which provides guidance in design, whether in terms of inspiration, basic  knowledge, or specific design issues, which have five sub-patterns in total. The other is the "Activist," which refers to to specific actions, including four sub-patterns: Spatial Layout, Basic Operation, Model Import, and Style Scheme. Regarding \textbf{Model Import}, we provided two options for the AI to deliver models: Model Retrieval, where the AI automatically searches from a large online model database for matches, and Model Generation, where the AI generates models directly. Interestingly, although our team hypothesized that more participants would prefer Model Retrieval due to its ability to deliver high-quality models that match users' basic needs quickly, more participants chose Model Generation. This may be due to curiosity about this emerging technology, which they hoped would inspire creativity. This phenomenon was observed multiple times during the experiment: participants repeatedly tried the Model Generation function, failing to achieve their desired results but eventually discovering that one of the generated models could work within another design concept. They then shifted their creative direction accordingly. Another notable pattern, \textbf{Style Scheme}, which is very important to users with a strong focus on visual style. It allows for the customization of style schemes, as well as the stylistic adjustment of individual models; it can even transfer existing style references into current schemes, altering colors, lighting, and model shapes of the existing arrangements.

\begin{table}[h]
  \caption{Categorization of Sub-patterns}
  \label{tab:Categorization}
  \centering
  \begin{tabular}{p{2cm} p{3cm} p{9cm}} % Adjust the widths as needed
    \toprule
    \textbf{Patterns} & \textbf{Sub-patterns} & \textbf{Description} \\
    \midrule
    \multirow{5}{*}{Guider} & Inspiration & This tool is often used in the initial stages of a project to generate inspiration. \\
    & Knowledge & A specialized knowledge base for spatial design where you can search for any relevant and useful information. \\
    & Refine Prompt & Refine your natural language to a sophisticated prompt. \\
    & Visual Reference & Generate Reference Visual Image of Scenes. \\
    & Review & Conduct reviews and scoring. Once a portion of your design is completed, you can use this tool to have the AI review your design and provide suggestions. \\
    \midrule
    \multirow{4}{*}{Activist} & Spatial Layout & This function is used for generating spatial layouts and projects them directly onto the ground. \\
    & Basic Operation & This function is used for helping users to operate models and other elements. \\
    & Model Import & Retrieve models by using natural language search from a large, custom-built database in the background, or generate high-fidelity models using text prompts, sketches, or low-fidelity models. \\
    & Style Scheme & Generate textures, form, color and light scheme for a specific model or the space. It can also be used to automatically adjust all selected models to maximize style consistency. \\
    \bottomrule
  \end{tabular}
\end{table}

Notably, the majority of participants (n=8) utilized a function within the I-T (Image to Text) category called Review. This function was particularly useful after Task 1, as it allowed AI to review the entire design by analyzing provided screenshots and offering suggestions for further refinement, guiding participants in completing the more detailed designs required in Task 2. While only one participant used the I-I (Image to Image) function, its potential, particularly in style transfer, was evident. The unexpected success of this participant's design highlights the function's ability to transfer a consistent style across all generated models, ensuring coherence in the overall scene's aesthetic.

Besides the above functions, the “ERROR” cards were also used several times, which revealed users' impractical expectations of AI capabilities, alongside gaps in their understanding of spatial design and modeling logic. These insights offer a deeper understanding of the novice-AI interaction: \textbf{1) Expectation vs. Capability Gaps for AI.} Many participants expected the AI to autonomously generate entire functional scenes or accurately interpret the modeling context, but the AI struggled with tasks that required more incremental design steps or detailed spatial awareness. For example, p2 requested an "accessible" cube space but received a solid cube model instead. P8 expected the AI to create a “cylinder table” within the existing scene, but the AI misinterpreted this and generated a generic operation desk. This overestimation reflects the need for clearer guidance on the limitations of AI in specific modeling contexts. \textbf{2) Spatial Design and Modeling Logic.} This is, users often lacked foundational spatial design knowledge, which contributed to errors. For example, p7 could not correctly apply a texture to a curved wall because they did not account for the complexities of UV mapping. This points to the need for guidance or tutorials that teach spatial design fundamentals alongside AI usage. \textbf{3) Reinterpreting User Requirements and AI Functions.} In some cases, users selected the wrong AI function for the task at hand, resulting in errors. For instance, when p12 tried to place trees in their scene, they called the model-generation AI, which could not handle the request. However, when the wizard switched to text-based AI and generated a guide (e.g., “place trees under the streetlights”), the user was able to quickly follow the instructions and improve their scene. This suggests that, at times, user requirements may need reinterpretation. The system could offer suggestions or alternative AI functions based on the task, helping users select the most effective tool for their design process.

\subsubsection{Common Workflow}

The interaction process of the 12 participants was distilled into a common workflow, which most participants followed, as illustrated in Figure~\ref{fig:workflow}. The process consists of four key stages:

\begin{figure}[h]
  \centering
  \includegraphics[width=\linewidth]{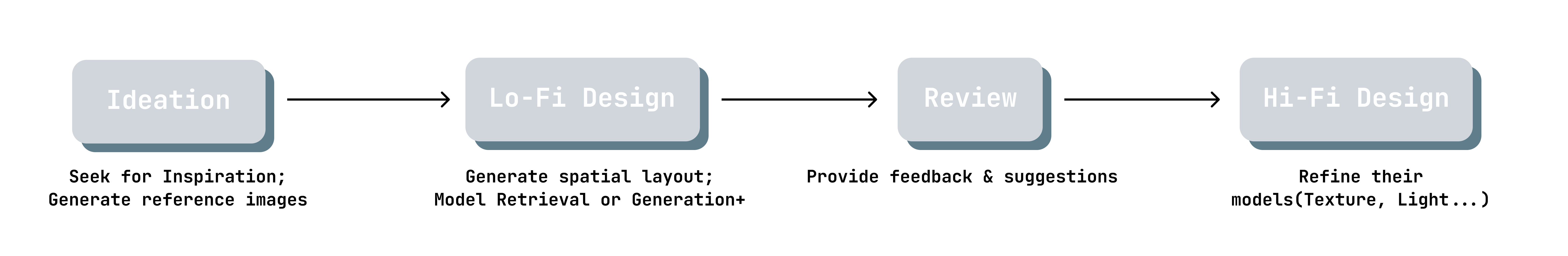}
  \caption{the Common Workflow: Ideation - Lo-fi design - Review - Hi-fi design}
  \label{fig:workflow}
  \Description{Common Workflow}
\end{figure}

\begin{itemize}
\item {\texttt{Ideation Phase: }}In the initial phase, participants typically used T-T (Text to Text) to seek inspiration from the AI, asking for general design advice for the type of space they were tasked with creating. The AI-generated descriptions were then refined into more specific prompts. Participants would then use T-I (Text to Image) to generate reference images, which served as the foundational style guide, helping them form a concrete design concept.
\item {\texttt{Lo-Fi Design Phase: }}In this phase, participants manually built low-fidelity models of the design, establishing key elements such as the position, size, and shape of individual components using basic geometric forms. Alternatively, some participants ask for AI assistance, using T-I (Text to Image) to generate spatial layout plans, which were imported and overlaid on the workspace to guide the construction of the low-fidelity models. This phase also involved the use of M-M (Model to Model) to convert each low-fidelity model into a high-fidelity version. In most cases, users did not complete all models, leaving the design in a draft state where low- and high-fidelity models coexisted.
\item {\texttt{Review Phase: }}In this stage, participants used I-T (Image to Text) to take screenshots of their partially completed design from specific perspectives and submit them to the AI for review. The AI then provided feedback and suggestions for further refinement, offering directions for enhancing the design.
\item {\texttt{Hi-Fi Design Phase: }}Based on the AI's feedback, users proceeded to refine their models. They used T-I (Text to Image) to generate textures for objects or employ I-I (Image to Image) to have the AI generate lighting schemes directly onto the image. 
\end{itemize}

Moreover, throughout both the Lo-Fi and Hi-Fi stages, users frequently sought AI assistance for operational tasks, such as helping to model a specific object or move an item to a particular position.

\begin{figure}[h]
  \centering
  \includegraphics[width=\linewidth]{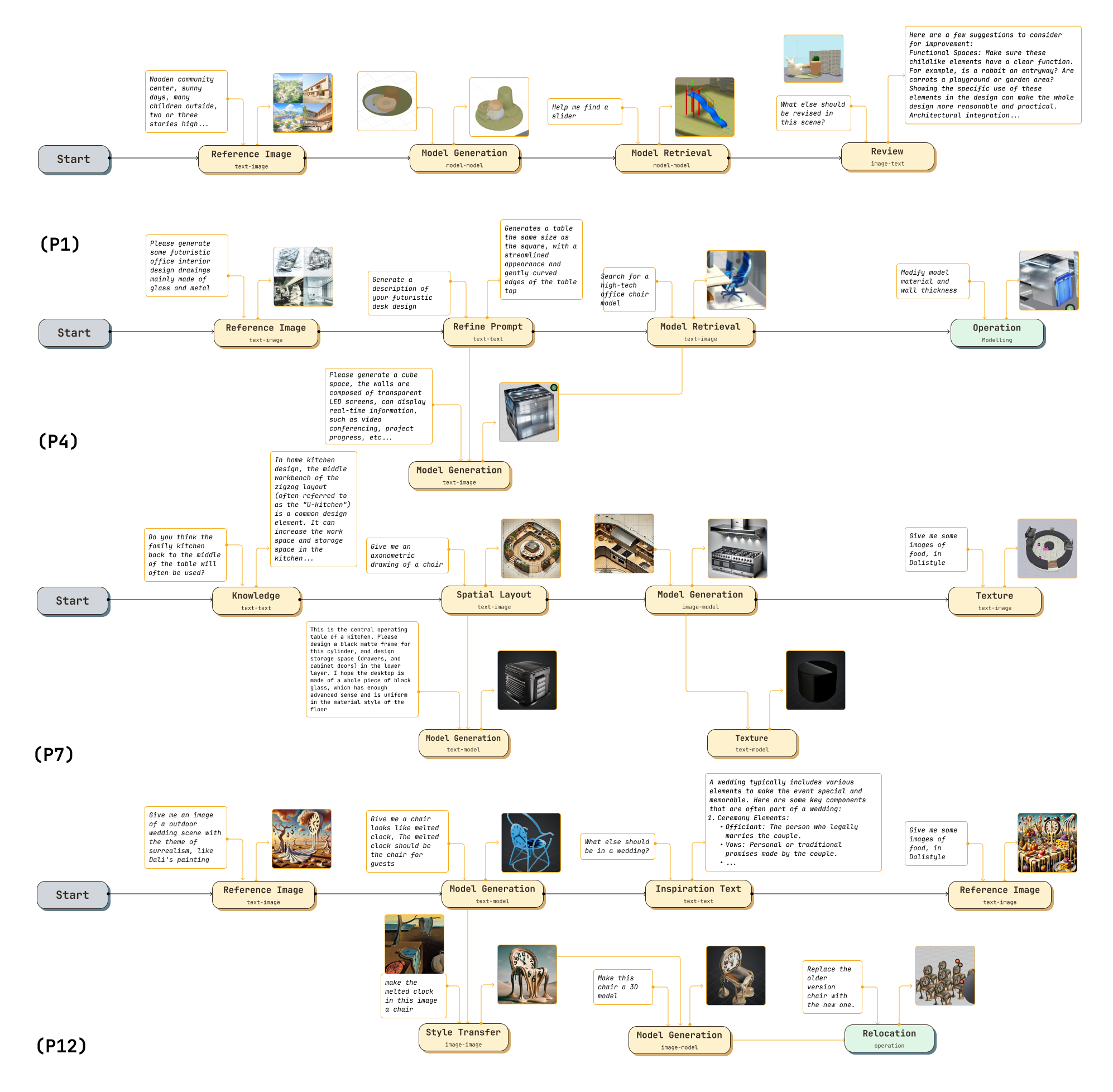}
  \caption{Specific workflow examples that have multiple differences compared with the common workflow: P1 jumped some steps; P4 seperated the generation of space and individual models; p7 generated models from an isometric view without using traditional layout plan; p12 is considered highly innovative by using style transfer to unify the visual styles without using spatial layout}
  \label{fig:examples}
  \Description{Specific workflow examples}
\end{figure}

While the workflow outlined above represents the common process followed by most participants, there were also unique and unexpected workflows observed, as depicted in Figure~\ref{fig:examples}. These alternative workflows were of particular interest, as they reflected the participants' varying levels of expertise, creative thinking, and personal approaches to design.

For example, some participants skipped certain phases of the process due to their prior experience or well-defined ideas. Participant P1, for instance, had some prior experience with spatial design and thus did not require AI inspiration during the ideation phase. Instead, she immediately used T-I (Text to Image) to generate reference images. Although she initially attempted to use AI to generate models, the results did not meet her expectations. However, she found that the AI-generated models had a distinct, concrete style that could be leveraged for the overall aesthetic of her design. This led her to repeatedly use the AI's model retrieval function, progressively adding concrete-style models to the scene and shaping the design accordingly.

Another interesting variation was observed in Participant P4, who approached the design task by separating the space and models. Designing a futuristic office scene, he treated the space as a model itself, first generating a textured "room box" for the environment and then incrementally generating individual models for doors, windows, chairs, and tables. This modular approach allowed him to focus on different elements of the design separately, creating a flexible and organized workflow.
Participant P7 employed a completely different approach by bypassing the need for a traditional floor plan. Instead, they instructed the AI to generate an isometric view, which simultaneously conveyed both the spatial layout and aesthetic style. The participant then extracted individual images from this isometric view and input them into the AI for further model generation, ensuring stylistic consistency across the generated models. Although this method posed some challenges due to current technological limitations, it showed great potential for maintaining coherence in design styles, and future advancements could further enhance this technique.

A particularly innovative approach was demonstrated by Participant P12, who utilized the I-I (Image to Image) function's style transfer feature—something no other participants explored. By applying the distinctive style of the surrealist master Salvador Dalí to the generated models, P12 achieved remarkable results. The workflow of this user is very suitable for those who care more about the visual styles.

\subsection{Thematic Analysis of Semi-structured Interview}

The thematic analysis is to flexibly capture the diversity and complexity of experiences of user experiences, aiming to uncover conclusions that may not have been anticipated under typical circumstances. The coding process follows the standard process\cite{41,49}. Two researchers coded the transcriptions independently, then discussed and iterated several times to organize the codes into themes. To ensure the accuracy and completeness of the data, all interviews were recorded and then manually transcribed verbatim by the research team members, and carefully reviewed multiple times to avoid any omissions or errors. Through the coding process, six themes were identified: 1) Quality Evaluation of AI-Assisted Functions; 2) Flexibility of Generated Content; 3) The characteristics of workflow; 4) The opportunities and challenges in communication; 5) The influence of embodied context; 6) Ways of AI to enhance or hinder your creativity and efficiency. These themes were further distilled into four main themes: Interaction Functionality, User Control and AI Autonomy, Novice-AI Communication, Creativity, and Efficiency.

\subsubsection{Interaction Functionality}

\textbf{Sub-theme 1.1: Quality Evaluation of AI-Assisted Functions}
This theme includes users' evaluation of AI-assisted functions and recommendations for function revisions based on their experiences during the design tasks. Participants generally rated the overall consistency and structure of the design tasks highly. They were particularly satisfied with the quality of image generation, text generation, and model retrieval functions. However, dissatisfaction was expressed regarding the quality of model generation, especially in complex design scenarios, where participants found the precision and customization of the generated results lacking. One participant noted: "AI-generated models in actual effect meet the request but are somewhat off from my expectations." Another commented: "The model is ugly, you see,  wrought iron shelves with country-style tiles. I wish the quality were better." Additionally, some participants raised concerns about the AI’s input prompt system, suggesting that many features within the AI's interface were underutilized. As one said, "The problem is with the prompts in AI; many features in the AI's nine-grid menu were not used."

Many participants provided suggestions for improving AI functions. The Review feature, which allows the AI to analyze and provide aesthetic feedback, was praised for helping users refine their designs. Participants also appreciated the AI-generated color schemes, which they found useful for enhancing visual appeal. Suggestions for further improvements also included the addition of a preset material library for easier access to textures and lighting schemes. Participants also expressed interest in enhanced customization options, such as personalized interface layouts and functional shortcuts tailored to different design workflows. As one participant suggested: "If shortcuts could be customized, it would be more convenient for me because different modeling software has different shortcuts, and I need to adapt to the new ones."

\subsubsection{User Control and AI Autonomy}

\textbf{Sub-theme 2.1: Flexibility of Generated Content.} This sub-theme focuses on the balance between user control and AI autonomy in the design process, particularly the flexibility of AI-generated content and the degree of user control over outcomes. Users desired more flexibility in the content generated by the AI. Several participants expressed a wish for the AI system to provide multiple design options, allowing users to iterate quickly and make adjustments if they were dissatisfied with the initial output. One participant noted: "The AI could generate multiple images for users to choose from and refine, similar to MidJourney; if users are not satisfied with one, they can explore other options." This reflects a broader sentiment among users: they prefer an AI system that provides varied options during content generation, rather than a one-size-fits-all approach.

In addition to these concerns, several participants criticized the "black-box" nature of AI systems, where the process behind the generation of content is not transparent to the user. One respondent remarked: "AI search is like a black box; you can’t interfere with the intermediate operations, while retrieval allows more control." This sentiment highlights the desire for more user involvement during the generation process, rather than simply accepting the AI's autonomous decisions.
 
\textbf{Sub-theme 2.2: The characteristics of workflow.} This sub-theme addresses the usability of the AI system, particularly in relation to workflow structure and guidance, especially for novice users. Several participants noted that a structured workflow might be more suitable for beginners, helping them navigate the system more easily. One user commented: "Having a fixed process feels more appropriate for beginners." Additionally, multiple respondents pointed out that the system lacked detailed operating guides and instructional documentation, which made learning the system more challenging. One participant stressed: "For beginners, detailed prompt documentation is essential so that I know how to describe the design I want." Another participant emphasized the importance of initial guidance before starting the tasks, suggesting: "Guidance is needed before the experiment; they could show examples from earlier participants."

\subsubsection{Novice-AI Communication}

\textbf{Sub-theme 3.1: The opportunities and challenges in communication.} Participants identified several communication challenges when interacting with the AI. Some noted that the AI often misunderstood their requests, particularly when converting images into models, which led to the generation of unwanted elements. One user remarked: "The AI had difficulty understanding exactly what kind of image or model I needed, and it struggled to grasp what needed modification."

When facing difficulties, some participants mentioned they were reluctant to communicate further with the AI. However others would refine their textual input or switch to a different image rather than attempting further clarification with the system.
 
\textbf{Sub-theme 3.2: The Influence of Embodied Context.} Many participants highlighted the limitations of AI in understanding the broader context of their design processes. One recurring issue was the AI's inability to track and reference prior interactions or actions. Users expressed frustration that the system lacked an embedded memory or history feature, which they felt would have significantly improved its ability to interpret their intentions based on previous design steps or adjustments. This sentiment was echoed by others, with one user explaining: "If the system could recall the last few models I worked on, it would make refining designs so much easier. Right now, it feels like the AI is always in a vacuum, unaware of what I’m trying to achieve overall." Participants also commented on the system’s struggle to maintain design consistency across multiple iterations. As one user pointed out: "I wanted the AI to generate a chair that fit with the table it made earlier, but it couldn’t 'see' the table. I had to describe everything again, and even then, the styles didn’t match." This lack of context-awareness led to frustration, especially when users had to repeatedly provide detailed descriptions or re-upload visual references.

Additionally, participants expressed a desire for the AI to better understand not just the static design elements, but also the evolving nature of their project. One user said: "As my design ideas evolve, I want the AI to adapt with me. If it could track the changes I make, it might offer suggestions that are more aligned with where I’m heading, rather than where I started."
 
\subsubsection{Creativity and Efficiency}
\textbf{Sub-theme 4.1: Ways of AI to enhance or hinder your creativity and efficiency.} This theme examines how AI influences creativity and efficiency during the design process, highlighting both the benefits and limitations of AI in enhancing user creativity. Participants reported that AI-generated images were often inspired when they had vague design concepts, helping them refine their ideas. One participant shared: "The AI-generated images gave me a lot of inspiration. At first, my ideas were unclear, but the images helped me develop them further." However, some users felt that while AI boosted their efficiency, it did not significantly enhance their creativity.

Interestingly, several participants noted that the AI sometimes guided them toward unexpected creative directions, which proved beneficial. For instance, one participant mentioned: "I wasn't happy with the initial AI suggestions, but then it led me in a completely new direction, which turned out to be useful and inspiring." Another respondent said: "AI's random generation feature helped me break out of traditional thinking patterns."

Participants also highlighted that AI-generated content is particularly well-suited for virtual scenarios, allowing for the creation of imaginative and unconventional designs. However, some users also pointed out that AI's output often lacked precision, which limited its utility in certain design contexts. As one participant noted: "Although AI-generated content is helpful, it sometimes lacks the precision I need to achieve my vision." This feedback underscores the need for further refinement in AI’s ability to balance creativity with user-specific design requirements.

\section{System Design Implications}
We propose several system design implications aimed at fostering a deeper understanding of the operation of the interactive system and presenting novel interaction methods in an intuitive and integrated way. Additionally, these implications seek to identify potential shortcomings in the system, thereby laying the groundwork for future development and user testing.

\subsection{Design Guidelines}

Based on the results of the research, we have established the following design guidelines:

\textbf{DG1: The tool should allow users to access it at any point during the design process and understand embodied commands based on the user’s current context.} We are interested in the progressive nature of the design process, meaning that users do not complete all operations with a single command. Instead, they should be able to call up the AI interface whenever needed over time, seeking help from the AI as necessary. To recognize embodied commands, we can clarify the context of the user's intent through screenshots or by using the mouse to select a specific model area for in-situ discussion.

\textbf{DG2: The Tool should provides multiple functions that balancing both creativity and effciency.} The balance between creativity and efficiency is crucial in the development of AI tools. On the one hand, AI needs to possess sufficient capabilities to provide immediate feedback when users make practical requests, ensuring the generated output is of appropriate quality. On the other hand, AI must retain a degree of flexibility to inspire human creativity. In this context, the implementation of multiple tools becomes necessary to fit with diverse needs.

\textbf{DG3: The tool should have an adaptive and integrated system that progressively adjusts the UI and user experience based on the user's level of expertise.} Through our experiments, we observed that novice users often feel overwhelmed by complex tasks when interacting with AI tools, especially in design contexts. To address this, the adaptive system helps by initially presenting simpler, essential functions. As users gain proficiency, the system gradually introduces more advanced features and controls. This structured, progressive design experience ensures that users aren’t overloaded with complexity early on but can still explore more advanced capabilities over time. Moreover, it is obvious that these tools should be well classified and integrated into the system, rather than scattered in different places. 

\subsection{Design Proposals}

\begin{figure}[h]
  \centering
  \includegraphics[width=\linewidth]{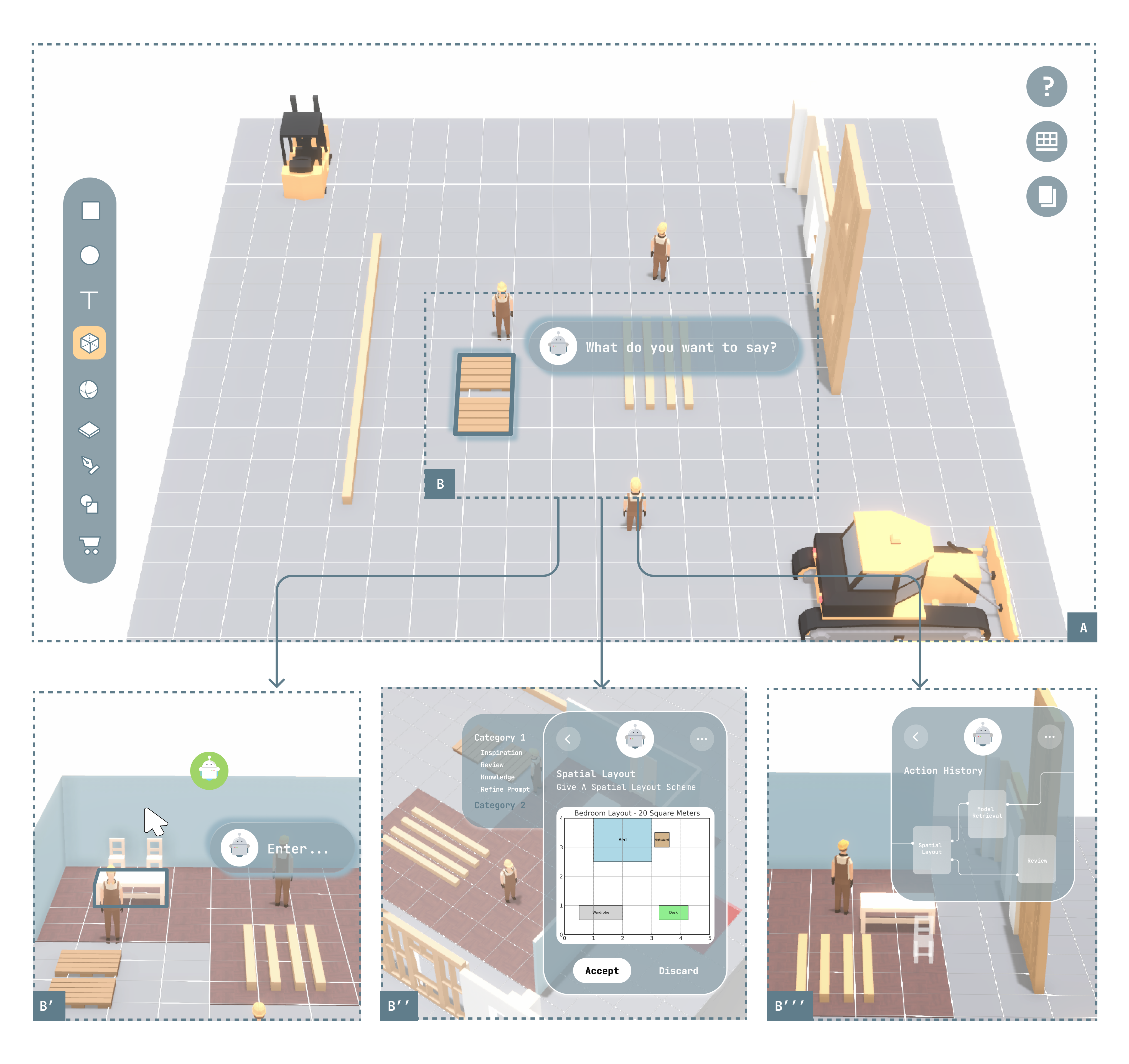}
  \caption{(A) the Main interface with multiple tools and the information panel; (B) Floating AI Spirit  with three states: B’, B’’, B’’’}
  \label{fig:interface}
  \Description{(A1) Main interface; (A2) Advanced 3D modeling tools; (A3) Object Information Panel; (B) Floating AI Spirit  with three states: B’, B’’, B’’’}
\end{figure}

As depicted in Figure~\ref{fig:interface}, the system consists of two main parts (A and B). The AI system (B) is built on top of the basic function panel (A), and it moves along with the position of your cursor, which can switch between states based on your commands:

The most significant change in Interface B is the integration of B4, a floating AI assistant. Unlike the fixed side panel in A5, this assistant appears as a floating "spirit" that follows the user's cursor. It accepts in-situ user commands in real time and translates them into visualized functional component logic. The AI assistant operates in three modes: \textbf{1) State 1 (B’) -- Embodied Spirit:} In this mode, the AI follows the user's cursor and can be hidden if desired. It operates in the context of either the entire window or selected models. When a conversation starts, the screen is locked, and a communication circle is left behind once the interaction ends. These circles can be toggled on or off in the settings. \textbf{2) State 2 (B’’) -- Card Selection Interface:} By clicking on the floating spirit in State 1, users can access a card-based interface with various predefined actions. If the project is new, the Inspiration panel is the default view, guiding users through the idea-generation process. \textbf{3) State 3 (B’’’) -- Operation History Interface:} This mode, accessible by clicking on the gear icon, allows users to review and adjust all past operations. The interface is presented as a visual logic diagram, where parameters can be modified, and the AI generation process can be restarted from a previous point.

In terms of functionality, firstly, various card selections provided in the AI system are configured based on patterns identified in prior research for diverse spatial design requirements. And in the tool panel in the main interface (A),  users can work with primitive shapes such as circles, rectangles, spheres, and cubes. Additionally, tools like the pen and Boolean operators are included for greater flexibility. As for the Store tool, it provides multiple modular and prefabricated models to choose and load into the scene.

Morever, in the context of modeling and the use of AI, we propose an adaptive system that offers different levels of functionality and entry points based on the user’s expertise. For beginners—those who are new to both modeling and AI—the system would initially provide simplified tools such as primitive building blocks, preset color palettes, and basic generative AI features, like text-to-image that generates a basic layout. As users grow more comfortable and proficient, they may seek to perform more complex tasks such as joining blocks, adding materials, or refining details. At this stage, the system would introduce additional features, like custom AI functions for adjusting textures or lighting. By first providing the most commonly used AI functions, the system ensures that novices aren't overwhelmed. As users’ needs evolve, they can prompt the system for new capabilities, and these advanced functions will be dynamically integrated into the current interface. This approach allows for a gradual learning curve and ensures that users can explore progressively more complex design tasks at their own pace.

\subsection{Technical Pipeline Testing}

We conducted an initial technical feasibility test to ensure that the system can be successfully built and operated in the future. The core of this system is based on the communication between Unity and Python. The front end is run by Unity, while the backend workflows and functionalities are executed by Python through external API calls. The generated assets are automatically imported into Unity via a communication interface.

Regarding 3D generation, we developed a minimal viable prototype (MVP) that can transform low-fidelity user models (abstract models composed of basic geometric shapes) into high-fidelity 3D models with textures. After the user creates a low-fidelity model and inputs a prompt, the system automatically captures a screenshot and sends it to the backend Python API. The system then utilizes ControlNet (to control the geometry outlines), SDXL (to generate high-quality renders), and TripoSR (to convert the renders into 3D models). We also independently tested other features, such as text-based dialogue, text-to-spatial layout generation, and text-to-image generation, all of which were successfully executed through API calls.  As for the modeling functions, they can be developed based on the Grid Building System, a grid-based modeling framework in Unity, which is particularly suitable for novice users.  

Overall, the prototype operates successfully, although the speed of generation, especially the model generation is relatively slow, which usually takes 3-5 minutes, accompanied by occasional lag. Improvements in the integration of functions and the running speed of AI models can be achieved through further technical optimization.
 
\section{Discussion}
\subsection{Contributions}
The core contribution of this article is an in-depth exploration and understanding of the progressive interaction mechanisms between novice users and AI in spatial design, achieved through a wizard-of-oz study. This experiment recorded multiple interactions between participants and wizards, identifying novice users' functional needs for AI, workflow patterns, and finding interaction opportunities and challenges through semi-structured interviews. System Design Implications provide intuitive demonstration and thinking for this progressive collaboration tool.

In the initial stage, we did not conduct formal experiments but built on pilot studies, which also reflects the theme of progressive design. The pilot study set the objectives for the formal experiments—observing behaviors and summarizing patterns. It also determined the tools for the experiment, including various AI-assisted tools, as well as the recording tool(function cards), which had not been fully considered before. The time and method for the experiment were also established in this phase: given the significant gap between the spatial design levels of the participants and the experimenters, it is hard for experimenters to predict the time and results of users. Moreover, interviews conducted after the study were crucial; their findings provided different but interesting perspectives that complemented the findings in the Wizard-of-Oz study. For example, the interviews pointed out the challenges of communicating with AI, but the workflows and patterns assumed smooth communication and did not consider the difficulties of the process. Some users chose to accept the flawed outputs from AI, altering their original design ideas; others refined their prompts for regeneration. 

The design proposal, which includes a novice-friendly modeling system and an embodied AI assistant, addresses a major gap in the current state of AI-assisted design tools by moving beyond one-step, uneditable solutions and introducing a progressive, iterative framework. This shift allows novice users to gain a deeper understanding of spatial design, while simultaneously boosting their creative abilities. It is crucial that three innovative design guidelines have been proposed, each corresponding directly to the discussion within the design proposal. Although a complete system has not yet been developed, these design guidelines significantly guide the direction for further refinement.

\subsection{Insights}
One key insight from our study is the gap between current AI development and the need for context-aware scene understanding. While current generative 3D AI tools focus on creating objects or models, they often do so without taking the broader environmental context into account. This disconnect between scene understanding and generative AI is evident in our use case, where we found that users struggle to provide control inputs to 3D AI systems that are consistent with the visual and conceptual elements of their environment. In our experiments, we noted that, for example, sometimes users needed the object to align with the broader stylistic and spatial characteristics of the scene, while in other cases, it could exist as an independent element. To accommodate these needs, we designed a context-selection function to make sure the AI understands the current context. Similarly, a related research \cite{3} has attempted to understand how humans prompt embodied commands to AI, but it focuses primarily on specific prompts on individual 3D operations rather than the systemic spatial design. Spatial design involves a collection of operations that cannot be determined by a single prompt alone. Therefore, it necessitates the establishment of a framework from a macro perspective, which can then be embodied to accommodate the complexities of spatial design.

Another key insight is the role of novice-AI workflow structure itself. We observed that for complete beginners, a fixed AI workflow yields more accurate results, as it provides clear step-by-step guidance. However, users with some prior experience exhibited a need for greater flexibility, particularly in fostering creativity. Results from pattern mapping indicate that while most users naturally follow a fixed process, a minority creatively utilize tools to achieve satisfactory designs.  To avoid segregating users based on their level of knowledge, we have adopted a flexible interaction approach. These findings point to the potential for developing an adaptive design system that dynamically adjusts its interface and features based on user needs and experience levels, supporting different creative processes and enabling more personalized and effective spatial design without adhering to a rigid modeling framework. Some Researchers have conducted similar work studying nonlinear Human-AI interactions based on creative design \cite{48}, which shares similarities with "progressive" interactions: both are based on discussions about the design process, positing that design is not a straightforward endeavor but a process of reflection, rejection, and iteration. This research also addressed the balance between creativity and accuracy, but the difference is, it relied on static categorizations of user expectations for interface adjustments. In contrast, our system is developmental: it functions like a level-based game, with more features and elements unlocked as the design process progresses. Users gain knowledge of spatial design through the design engagement with the tool, demonstrating its potential utility in spatial design education.

We have also observed that due to the complexity of spatial design, novices frequently encounter difficulties at various stages of the process. While the exact points of difficulty differ based on individual backgrounds, all users consistently required assistance from AI models at different stages. This observation prompted us to design an integrated system Unlike existing solutions that rely on scattered AI tools, our system combines multiple AI models into a unified interface, allowing users to seamlessly access the appropriate model when needed. Additionally, the framework includes a history browsing feature, enabling users to revisit and modify previous design iterations, which was rooted in our goal of supporting an iterative, flexible workflow where users can refine their designs continuously without having to start over. In terms of its interface design, the current choice is a visual logic graph, rather than merely a recorded collection of functional cards. This approach helps to visually represent the entire process of user design and allows for nonlinear outcomes of design evolution. It enables users to revisit previously failed stages while still being able to see the logical relationships between actions. Currently, some AI tools, such as ComfyUI, are adopting this approach, but ComfyUI is non-embodied and does not support the generation of spatial designs. Traditional parametric modeling tools like Grasshopper serve as embedded visual design logic in 3D modeling environments, but they are clearly insufficient in terms of AI integration. Additionally, based on our statistical analysis, we are developing the most commonly used functions and some potentially useful ones, while selectively excluding inefficient or unused features for integration into the existing framework.

\subsection{Limitations}
Our initial evaluation of the experimental results shows a satisfactory outcome. Most participants were able to complete the design tasks within the allotted time, with some surpassing expectations. However, it also revealed several limitations that need to be addressed in future iterations of the system. First, the "Wizard of Oz" method, used to simulate AI behavior, exposed inconsistencies in the feedback provided by human experimenters acting as AI agents. These inconsistencies stemmed from the fact that the experimenters had not undergone formal training to accurately mimic AI behavior. As a result, the quality and timing of AI feedback were sometimes less than ideal, affecting the participants’ experience and the flow of the design process. To improve this, future experiments should involve more comprehensive training for experimenters.

Second, technology. One is the model generation time from lo-fi to hi-fi model, which was longer than expected, taking 3-5 minutes to process. It is primarily caused by the complexity of the workflow.  Recent advancements, such as the release of TripoSR, have demonstrated the ability to generate models in as little as half a second on high-end GPUs like the A100.  However, these results are achieved through a streamlined, single-step workflow and require powerful hardware, which is not accessible to the average user. Therefore, optimizing the workflow to balance speed and functionality, while ensuring that it remains viable on consumer-grade hardware, is a challenge that requires further research and development. Another limitation lies in the AI's current ability to manage detailed and precise design tasks.  While the AI was effective at generating creative ideas and initial design concepts, it struggled with refining those ideas into high-fidelity spatial designs.  This challenge stems from the lack of AI models specifically trained for spatial design.

It can be observed that some of the limitations identified can be addressed with our advancements, while others
require further development in AI technology. In the short term, we will conduct rigorous training for AI experimenters,
refine the experimental process, and test with a larger pool of participants to gather more meaningful insights. Interface
design improvements will progress in tandem with system development, with a focus on streamlining workflows
and creating more intuitive interactions to enhance usability for novices. On the hardware side, improving hardware
integration is achievable, while on the software side, integrating the front and back end more effectively will help
reduce system lag and long generation times. In the longer term, addressing the challenges related to detailed and
precise spatial design tasks will require significant advancements in AI technology. This includes developing faster and more accurate AI models, specifically customized to spatial design that can manage complex layouts, textures, and spatial relationships.

\section{Conclusion and Future Work}
This study introduced an integrated, progressive system for novice-AI collaboration in spatial design, addressing the challenges novices face with scattered, one-step, and uneditable AI tools. By consolidating AI models into a unified framework with contextual flexibility, progressive development, and explainability, users can gradually advance their spatial designs based on context and make adjustments to previous stages. It employed a Wizard-of-Oz approach, utilizing functional cards to record the interaction workflows between novices and AI, thereby gaining a deep understanding of user behavior patterns and summarizing common functions and workflows. In addition, our interface design redefines the interaction paradigms of existing AI tools by introducing an innovative "floating spirit" mechanism, which integrates adaptive, embodied and retraceable characteristics, allowing for a more dynamic and flexible interaction. These contributions advance our understanding of Novice-AI interaction mechanisms and build a strong foundation for the development of AI-assisted spatial design tools specifically tailored for novice users.

While our experiments have identified user habits and needs, the full system is not yet fully developed. However, because of the successful completion of technical pipeline checks, we have validated the theoretical feasibility of the system’s architecture. Due to the system’s complexity and the technical challenges we encountered—such as long generation times and the absence of specialized spatial design models—further development has been delayed. Future work will focus on optimizing system performance, reducing processing times, and developing AI models specifically tailored to spatial design tasks, to develop a fully workable system. Once the system is finished, multiple rounds of user testing will be conducted to enhance its precision and usability, ultimately enabling the AI to provide better support to novice designers.

%%
%% The next two lines define the bibliography style to be used, and
%% the bibliography file.
\bibliographystyle{ACM-Reference-Format}
%\bibliography{sample-base}

\end{document}